# Very large Dzyaloshinskii-Moriya interaction in two-dimensional Janus manganese dichalcogenides and its application to realize skyrmion states

Jinghua Liang[1], Weiwei Wang[2], Haifeng Du[2], Ali Hallal[3], Karin Garcia[4], Mairbek Chshiev[3], Albert Fert[4,5], Hongxin Yang[1,6*]

[1] *Ningbo Institute of Materials Technology and Engineering, Chinese Academy of Sciences, Ningbo 315201, China*

[2] *Anhui Province Key Laboratory of Condensed Matter Physics at Extreme Conditions, High Magnetic Field Laboratory, Chinese Academy of Sciences and University of Science and Technology of China, Hefei 230026, China*

[3] *Univ. Grenoble Alpes, CEA, CNRS, Spintec, 38000, Grenoble, France*

[4] *DIPC and University of the Basque Country, 2018, San Sebastian, Spain*

[5] *Unité Mixte de Physique, CNRS, Thales, University Paris-Sud, University Paris-Saclay, 91767 Palaiseau, France*

[6] *Center of Materials Science and Optoelectronics Engineering, University of Chinese Academy of Sciences, Beijing 100049, China*

The Dzyaloshinskii-Moriya interaction (DMI), which only exists in noncentrosymmetric systems, is responsible for the formation of exotic chiral magnetic states. The absence of DMI in most two-dimensional (2D) magnetic materials is due to their intrinsic inversion symmetry. Here, using first-principles calculations, we demonstrate that significant DMI can be obtained in a series of Janus monolayers of manganese dichalcogenides MnXY (X/Y = S, Se, Te, X ≠ Y) in which the difference between X and Y on the opposites sides of Mn breaks the inversion symmetry. In particular, the DMI amplitudes of MnSeTe and MnSTe are comparable to those of state-of-the-art ferromagnet/heavy metal (FM/HM) heterostructures. In addition, by performing Monte Carlo simulations, we find that at low temperatures the ground states of the MnSeTe and MnSTe monolayers can transform from ferromagnetic states with worm-like magnetic domains into the skyrmion states by applying external magnetic field. At increasing temperature, the skyrmion states starts fluctuating above 50 K before an evolution to a completely disordered structure at higher temperature. The present results pave the way for new device concepts utilizing chiral magnetic structures in specially designed 2D ferromagnetic materials.

---

# I. INTRODUCTION

Chiral magnetic structures such as chiral domain walls [1, 2], helical structures [3, 4] and magnetic skyrmions [5, 6, 7] hold promise for potential applications in future spintronic devices. Microscopically, the Dzyaloshinskii-Moriya interaction (DMI), which favors canted spin configurations, plays an essential role in the formation of such noncollinear magnetic nanostructures. For the presence of DMI, in addition to strong spin-orbit coupling (SOC) and magnetism, the system is required to have broken inversion symmetry [8, 9]. Therefore, significant DMI typically arises in noncentrosymmetric bulk magnets [10, 11, 12] and at interfaces [13, 14] between a ferromagnet and an adjacent layer with strong SOC. Notably, much effort has been devoted to grow multilayer stacks of ferromagnet/heavy metal (FM/HM) heterostructures, e.g., Ir/Co/Pt [15], Ir/Fe/Co/Pt [16] and Pt/Co/MgO [17] multilayers, in order to enhance the interfacial DMI [18].

Two-dimensional (2D) materials are fascinating class of materials combining an extremely small thickness with novel physical properties related to their 2D character [19]. Recently, the experimental progress has led to a breakthrough in the synthesis of the long-sought 2D magnets, which are first realized in the $Cr_2Ge_2Te_6$ bilayers [20] and $CrI_3$ monolayers [21] at low temperature, and then in the monolayers of $VSe_2$ [22], $MnSe_2$ [23] and $Fe_3GeTe_2$ [24] around room temperature. The discoveries of these truly 2D magnets have opened up new opportunities for spintronic technology. However, most of the above 2D magnets are centrosymmetric such that the DMI is absent in these structures. It is possible to break the inversion symmetry by modifying the substrate, combining different 2D materials, applying a bias voltage or strain [25, 26]. However, the simplest and most desirable situation should be a 2D magnet with inherent inversion asymmetry, intrinsic DMI and intrinsic chiral textures.

Recent experiments have demonstrated that Janus monolayers of transition metal dichalcogenides (TMD), e.g. MoSSe [27, 28], can be synthesized by controlling the reaction conditions. The intrinsically broken inversion symmetry, together with the feasibility of tunable electronic properties by a selection of a suitable pair of chalcogen elements [29, 30, 31] in the Janus TMD monolayers, inspired us to speculate that large DMI can be obtained in the 2D magnetic Janus materials. With this conjecture, we investigate the structural and magnetic behavior of a series of Janus monolayers of manganese dichalcogenides MnXY (X/Y = S, Se,

Te, X≠Y) *via* first-principles calculations. Strikingly, we find that the DMI in MnSeTe and MnSTe monolayers is as strong as those in state-of-the-art FM/HM heterostructures. Furthermore, we apply Monte Carlo (MC) simulations to show that magnetic skyrmions can be stabilized in these 2D magnets.

## II. COMPUTATIONAL METHODS

Our first-principles calculations are performed within the framework of density functional theory (DFT) as implemented in the Vienna *ab-initio* simulation package (VASP) [32]. The electron-core interaction is described by the projected augmented wave (PAW) method [33, 34, 35]. The exchange correlation effects are treated with the generalized gradient approximation (GGA) of Perdew-Burke-Ernzerhof (PBE) [36]. In order to describe well the 3$d$ electrons, we employ the GGA+$U$ method [37] with an effective $U = 2$ eV for Mn as reported in the previous studies [38, 39]. The energy cutoff for plane wave expansion is set to 520 eV, and a $\Gamma$-centered 22×22×1 $\boldsymbol{k}$-point mesh is adopted for the Brillouin zone integration. All the structures are fully relaxed until the force acting on each atom is less than 0.001 eV/Å. Phonon dispersions are calculated with a 4×4×1 supercell by using the PHONOPY code [40, 41] along with the density-functional perturbation theory (DFPT) [42]. We have used the chirality-dependent total energy difference approach to obtain the DMI strength, which has been successfully employed for the DMI calculations in frustrated bulk systems and insulating chiral-lattice magnets [43, 44] and adapted to the case of thin films [45, 46]. In the calculations of DMI, a $\Gamma$-centered 20×5×1 $\boldsymbol{k}$-point mesh is adopted.

Using the magnetic interaction parameters determined by the first-principles calculations, we apply MC simulations with the Metropolis algorithm to explore the magnetic states. The investigated systems are gradually cooled down from 1000 K to the required low temperature. For each temperature, 2×10$^5$ MC steps are employed to thermalize the system. In all MC simulations, a large supercell of 160×160×1 unit cells with periodic boundary conditions is used in order to avoid the non-universal effects of boundary conditions. All MC simulations are performed with our JuMag package [47].

### III. RESULTS AND DISCUSSION

#### A. Geometric properties and structural stability

Figs. 1(a) and 1(b) show the top and side views of the crystal structure of the MnXY (X/Y = S, Se, Te, X≠Y) monolayers. One can see that the Mn atoms with point group $C_{3v}$ form a hexagonal network sandwiched by two atomic planes of different chalcogen atoms. Note that in our calculations, for the chemical formula MnXY, the lighter chalcogen atom X is always set in the top layer while the heavier atom Y in the bottom layer. The calculations of phonon spectrum (see Figs.1(c)-(e)) indicate that except for MnSSe, the other two monolayers are dynamically stable. For MnSSe, the small negative frequency in out-of-plane acoustic (ZA) phonon branch around the Γ point (see Fig. 1(e)) is related to the structural instability due to in-plane bending of the two different chalcogen atomic (S/Se) planes, which has also been reported in some nonmagnetic Janus monolayers [29]. The relaxed structural parameters of MnXY monolayers including lattice constants $a$, bond lengths of Mn-X(Y), $d_{X(Y)}$, and tilting angles of atomic planes Mn-X(Y)-Mn, $\theta_{X(Y)}$, are listed in Table I. The lattice constants of MnXY decrease as a function of the sum of X and Y atomic radii, and for a given MnXY monolayer, $d_X$ and $\theta_X$ are always smaller than $d_Y$ and $\theta_Y$, respectively, due to the smaller atomic radius of X atom compared to that of Y. Similar relationships of structural parameters are also found in the nonmagnetic Janus TMD monolayers [29, 30, 31]. The asymmetry between the top and bottom layers in MnXY materials breaks the inversion symmetry, thus allowing the DMI between the Mn ions, as we demonstrate in the following discussions.

Table I. The optimized lattice constants $a$, bond lengths of Mn-X/Y $d_{X(Y)}$, and tilting angles of atom planes Mn-X/Y-Mn $\theta_{X(Y)}$ of MnXY monolayers.

| Pattern | $a$ (Å) | $d_X$ (Å) | $d_Y$ (Å) | $\theta_X$ (deg) | $\theta_Y$ (deg) |
|---------|---------|-----------|-----------|------------------|------------------|
| MnSeTe  | 3.68    | 2.52      | 2.74      | 51.88            | 58.53            |
| MnSTe   | 3.60    | 2.38      | 2.74      | 48.31            | 59.90            |
| MnSSe   | 3.50    | 2.38      | 2.50      | 51.15            | 55.64            |

#### B. Spin model and magnetic parameters

In order to explore the magnetic properties of Janus MnXY monolayers, we fit the following model Hamiltonian for the spins of Mn atoms in the hexagonal structure:

$$H = -\sum_{<i,j>} \boldsymbol{D}_{ij} \cdot (\boldsymbol{S}_i \times \boldsymbol{S}_j) - J \sum_{<i,j>} \boldsymbol{S}_i \cdot \boldsymbol{S}_j - \lambda \sum_{<i,j>} S_i^z S_j^z - K \sum_i (S_i^z)^2 - \mu_{Mn} B \sum_i S_i^z \quad (1)$$

with the results of our DFT calculations. In Eq. (1) $\boldsymbol{S}_i$ is a three-dimensional unit vector representing the orientation of the spin of the $i$th Mn atom, and $<i,j>$ refers to nearest neighbor Mn atom pairs. The first three magnetic interaction terms including the DMI, the Heisenberg isotropic exchange, the anisotropic symmetric exchange and the easy axis single ion anisotropy are characterized by the parameters $\boldsymbol{D}_{ij}$, $J$, $\lambda$ and $K$ in the corresponding terms. The last term is the Zeeman interaction, where $\mu_{Mn}$ and $B$ represent the magnetic moment of the Mn atoms and the external magnetic field, respectively.

We first discuss the DMI, which is the most interesting parameter for this work. According to Moriya's symmetry rules [9], since the reflection planes pass through the middle of the bonds between two adjacent Mn atoms, the DMI vector $\boldsymbol{D}_{ij}$ for each pair of nearest neighbor Mn atoms is perpendicular to their bonds. Thus $\boldsymbol{D}_{ij}$ can be expressed as $\boldsymbol{D}_{ij} = d_{//}(\hat{\boldsymbol{u}}_{ij} \times \hat{\boldsymbol{z}}) + d_{ij,z}\hat{\boldsymbol{z}}$ with $\hat{\boldsymbol{u}}_{ij}$ being the unit vector between sites $i$ and $j$ and $\hat{\boldsymbol{z}}$ indicating normal to the plane. The in-plane component $d_{//}$ along with the associated SOC energy $\Delta E_{soc}$ can be evaluated by the chirality-dependent total energy difference approach [45, 46] with the two spin configurations depicted by yellow arrows in Fig. 1(b). Here we adopt the sign convention such that $d_{//} < 0$ ($d_{//} > 0$) favors spin canting with clockwise (counterclockwise) chirality. To calculate the out-of-plane component $d_{ij,z}$, we can use the relation $d_{ij,z} \approx d_{//}/\tan\tilde{\theta}_{ij}$ [26], where $\tilde{\theta}_{ij} = (\theta_{ij,X} + \theta_{ij,Y})/2$ represents the average of the tilting angle of atomic plane (Mn)$_i$-X(Y)-(Mn)$_j$. Note that although we include $d_{ij,z}$ in the following MC simulations, it does not play any dominant role since the sign of $d_{ij,z}$ changes in a staggered way for the six nearest neighbors of the Mn atoms, leading to a vanishing $d_{ij,z}$ in average.

Fig. 2 presents the calculated in-plane DMI component $d_{//}$ of Janus MnXY monolayers with the DMI vectors $\boldsymbol{D}_{ij}$ between the nearest neighbors of Mn atoms schematically shown in the inset. It is remarkable that all Janus MnXY monolayers have strong DMIs, especially for MnSeTe and MnSTe whose magnitude of $d_{//}$ reaches 2.14 meV and 2.63 meV, respectively. These values are comparable to many state-of-the-art FM/HM heterostructures, e.g., Co/Pt (~3.0 meV) [17, 45] and Fe/Ir(111) (~1.7 meV) [48] thin films that serves as prototype multilayer systems to host skyrmion states. Even for MnSSe with the weakest DMI among the investigated 2D magnets, the magnitude of $d_{//}$ (0.39 meV) is larger than that of the graphene/Co system (~0.16 meV), in which the DMI induced chiral domain walls have been reported recently [46].

To elucidate the origin of the exceptional DMI in Janus MnXY monolayers, we plot their associated SOC energy $\Delta E_{soc}$ in Fig. 3. One can see that in all MnXY monolayers the dominant contribution to DMI stems from the adjacent heavy Y atom, especially the heavy Te atom in MnSeTe and MnSTe that makes their DMI magnitudes much larger than that of MnSSe. Similar behavior has been identified for the FM/HM heterostructures [45], where $\Delta E_{soc}$ is dominated by the heavy 5$d$ transition metal at the interfacial layer. This is so called Fert-Levy mechanism of DMI which can be understood by considering that the heavy chalcogen atoms (5$d$ transition metals in the FM/HM heterostructures) act as spin-orbit active sites to induce the spin-orbit scattering necessary for the DMI [9, 49].

In table II, we summarize the remaining magnetic interaction parameters of the exchange coupling $J$, the anisotropic symmetric exchange $\lambda$ and the easy axis single ion anisotropy $K$ as calculated with the method illustrated in the supplemental material [50], and the magnetic moments $\mu_{Mn}$ of Mn atoms. All MnXY monolayers are ferromagnetic with $J > 0$ and have an out-of-plane easy axis with both $\lambda$ and $K$ being positive. Moreover, the calculated Curie temperature $T_c$ (see Table S1 in [50]) of MnSeTe, MnSTe and MnSSe is 170 K, 140 K and 190 K, respectively. Interestingly, previous researches [51] and our calculations show that tensile strain can enhance the exchange interactions in the manganese dichalcogenides monolayers, which indicate that $T_c$ of Janus MnXY monolayers can be increased close to the room temperature by appropriate change of system parameters (Fig. S2 in Ref. [50]).

Table II. The calculated parameters of exchange coupling $J$, anisotropic symmetric exchange $\lambda$ and easy axis single ion anisotropy $K$, and the magnetic moments $\mu_{Mn}$ of Mn atoms.

| Pattern | $J$ (meV) | $\lambda$ (meV) | $K$ (meV) | $\mu_{Mn}$ ($\mu_B$) |
|---------|-----------|-----------------|-----------|----------------------|
| MnSeTe  | 13.26     | 0.16            | 0.37      | 3.68                 |
| MnSTe   | 10.52     | 0.004           | 0.29      | 3.64                 |
| MnSSe   | 15.60     | 0.12            | 0.07      | 3.42                 |

### C. Spin textures from Monte Carlo simulations

Once all the parameters in the spin Hamiltonian are determined, we can perform MC simulations starting from an initial disordered state at 1000 K to check whether it is possible to stabilize chiral spin textures in the MnXY monolayers. Moreover, we note that while the DMI/exchange ratios $|d_{///}/J|$ of MnSSe is only about 0.03, those of MnSeTe and MnSTe can reach 0.16 and 0.25, respectively, which are within or larger than the typical range of 0.1-0.2 for the formation of skyrmions [52]. We thus only consider MnSeTe and MnSTe monolayers in the following discussions.

Our simulations immediately reveal that, in both systems, at low temperature (10 K) and in zero field, we obtain a ferromagnetic state with worm-like domains separated by chiral Néel domain walls (DW), the thin white lines between domains of up and down magnetizations in Figs. 4(a) and 4(d). Notably, the size of the domain in MnSeTe is much larger than in MnSTe, which is consistent with the smaller DMI/exchange ratio, $|d_{///}/J|$, and the resulting larger DW energy in MnSeTe compared with MnSTe. We have also found that different shapes and different orientations of the domains are obtained with different initial random spin configurations.

More importantly, next we find that, in both MnSeTe and MnSTe, skyrmion states can be induced by external magnetic fields as in many magnetic systems with skyrmions [15, 16, 17]. For the MnSeTe monolyer, applying a field shrinks the red domains (see Fig. 4(a)) and isolated skyrmions begin to appear at about 0.05 T. The worm domains completely disappear above 0.15 T and we show in Fig. 4(b) a typical image of the disordered assembly of skyrmions in a ferromagnetic background observed between 0.2 and 0.4 T. Above 0.4 T, the density of skyrmions decreases and a uniform ferromagnetic state without skyrmions is set above 0.6 T. The diameters of the skyrmions (diameter of the white circle with in-plane magnetization) decreases from 6.6 to 5 nm between 0.2 and 0.55 T.

For MnSTe, we find a similar tendency evolving from a ferromagnetic state with worm domains at low field (see Fig. 4(d)), to **a lattice of** of skyrmions for fields in the 1.4-1.8 T range (see Fig. 4(e)). The density of skyrmions increases as the field increases from 1.4 to 1.8 T and they finally tend to form an approximate hexagonal lattice. Their diameter decreases from about 8 nm to 7 nm at increasing field. The crossover from individual skyrmions in MnSeTe to skyrmion lattice [53] is consistent with the larger DMI in MnSTe.

As the temperature increases, the images of the skyrmions in MnSeTe and MnSTe begin to be less well-defined above about 50K and become more and more blurred, as shown on Figs. 4(c) and 4(f) for $T = 150$ K. This blurring expresses the destabilization of the skyrmions by thermal fluctuations. To obtain a more quantitative information on the thermal destabilization of the skyrmions, we have derived from our simulations the temperature and field dependence of topological charge, $Q = \frac{1}{4\pi} \int \boldsymbol{S} \cdot (\partial_x \boldsymbol{S} \times \partial_y \boldsymbol{S}) dx dy$ [54, 55], of MnSTe per supercell. The skyrmion lattice phase of MnSeTe corresponds to the blue area shown in the phase diagram of Fig. 5. Here an external field $B$ pointing upwards along the out-of-plane direction leads to skyrmions with $Q = -1$. One can clearly see from Fig. 5 that, approximately above 1 T and below 125 K, there is a large blue area associated with significant negative $Q$ in the $B$-$T$ plane, which signals the formation of the skyrmion states (**26 in Fig. 4(e) corresponding to Q = -26).** We also find that at this area, for a fixed $B$, the number of skyrmions in the system does not change significantly as the temperature is increased up to 90 K. When the temperature is further increased, the magnitude of $Q$ starts decreasing and eventually goes to zero. The decrease of the magnitude of $Q$ is associated with a fluctuation-disordered state (see Fig. 4(f) for $B = 0.3$ T and $T = 150$ K), where the skyrmion lifetime is finite. Similar relations between thermal fluctuation of skyrmions and the evolution of the density of topological charge were already discussed in [56, 57]. We note that the temperature dependence of magnetic parameters [58] can also affect the thermal stability of skyrmion at finite temperature. Since the calculation of temperature dependence of exchange interactions is outside the scope of the present study, we would leave it for future research.

The most interesting skyrmions for applications are the individual skyrmions which can be manipulated individually in devices such as racetrack memories or logic components [59]. Such individual skyrmions are those shown in Fig.4b for MnSeTe. The field range (0.2-0.4 T) in which they exist in MnSeTe is somewhat higher than the typical range (0.02-0.08 T) found in metallic multilayers [59]. It can be a disadvantage for applications. A possible way for the reduction of the DMI and the corresponding reduction of the field scale in MnSeTe can be use of proximity effects by integration in van der Waals heterostructures. In addition, there are several alternative approaches to replace an applied field for the stabilization of skyrmions. For instance, one can use exchange biasing with antiferromagnet as proposed recently [60] and demonstrated for the stabilization of skyrmions in zero field. Exchange-biasing with a ferromagnetic layer has also been demonstrated [61].

## IV. CONCLUSION

In summary, using first-principles calculations we demonstrated that strong DMI can be obtained in Janus MnXY monolayers with inherent inversion asymmetry. We find that the strong DMI stems from the large SOC in the heavy chalcogen atoms. The MC simulations show that, in the low field limit, MnSeTe and MnSTe host worm like domains separated by chiral Néel domain wall. Moreover, external field can break the worm like domains and create skyrmion states in these two monolayer structures.

As compared with FM/HM multilayers that are fabricated with stacks of different materials to enhance the DMI and generate skyrmions, using a single 2D material as the MnXY monolayers is a simpler situation and can probably be beneficial for a low density of defects. Moreover, our MnXY monolayers can be integrated in van der Waals heterostructures [19, 62] to extend their properties. One also knows from theoretical calculations by Gmitra *et al*. [63] that the proximity of graphene with a TMD can induce a large spin-orbit splitting of the Dirac cone of graphene with creation of Rashba-like Fermi contours. For the current-induced motion of skyrmions in MnXY, this gives the possibility of injecting a spin current into the MnXY layers by using the Edelstein Effect [64] of the Rashba-like electrons in graphene. Altogether, our calculation results suggest that the Janus MnXY monolayers are good candidates for spintronic nanomaterials and nanodevices.

*Note added*: After the submission of this paper, we noted that a recent work by Xu *et al*. [65]

predicted that the Janus $Cr(I, X)_3$ (X = Br, I) monolayers can show strong DMI and stabilize metastable skyrmions.

**Acknowledgment**

We thank O. Boulle and L. Buda-Prejbeanu for helpful discussions and acknowledge financial support from the National Natural Science Foundation of China (11874059), Zhejiang Province Natural Science Foundation of China (LR19A040002), and Key Research Program of Frontier Sciences, CAS, Grant NO. ZDBS-LY-7021. We also acknowledge the support from DARPA TEE program, DIPC and University of the Basque Country and Horizon 2020 Research and Innovation Programme under grant agreement no. 785219 (Graphene Flagship).

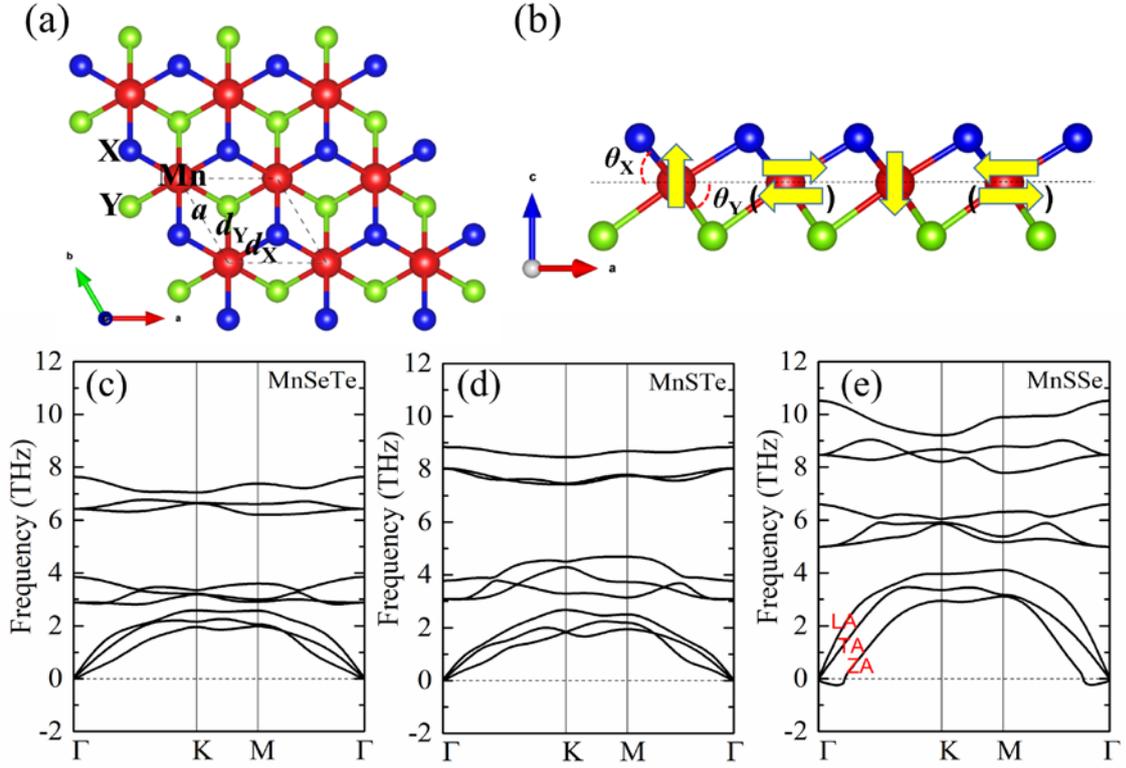

**Figure 1.** The top (**a**) and side (**b**) views of the crystal structure and the phonon spectrum (**c**)-(**e**) for the MnXY (X/Y = S, Se, Te, X≠Y) monolayers. The dash lines in (a) shows the primitive cell. The yellow vectors in (b) indicate the two spin configurations with opposite chirality used to extract the in-plane DMI parameters.

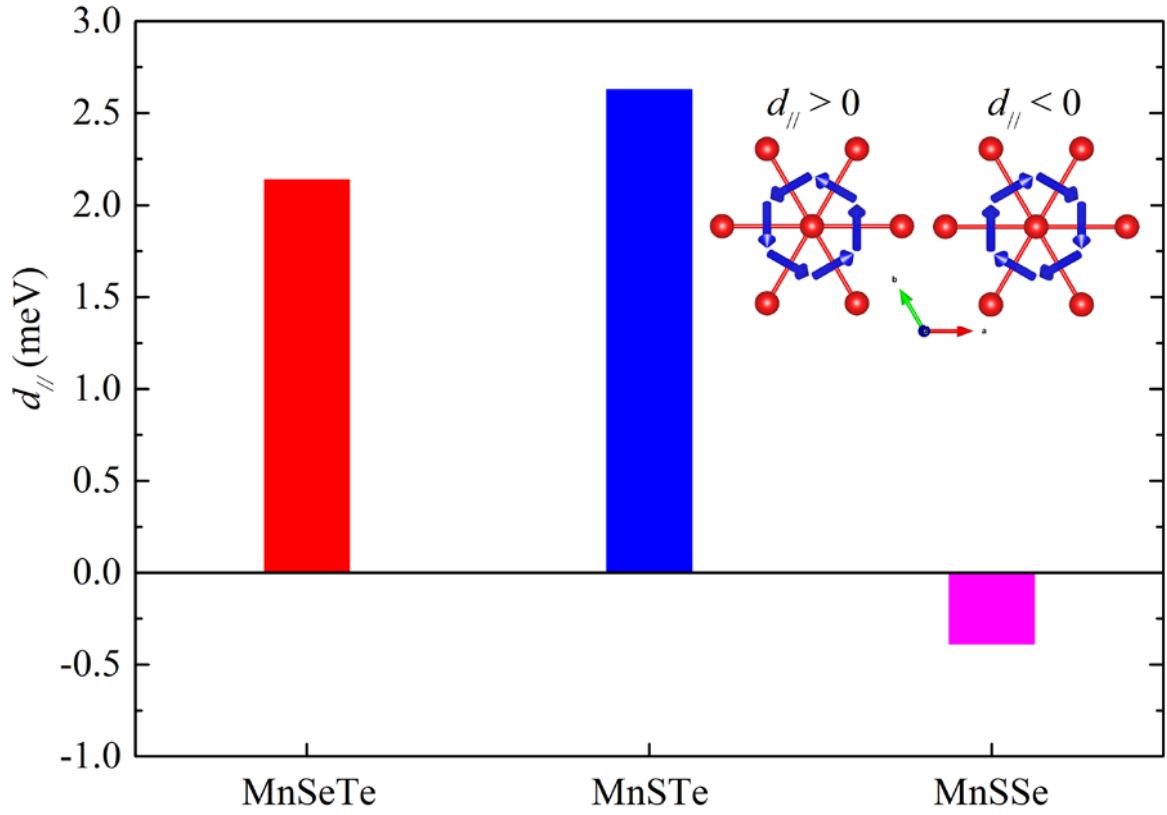

**Figure 2.** The calculated in-plane DMI parameters $d_{//}$ of the Janus MnXY monolayers. Here $d_{//}$ < 0 ($d_{//}$ > 0) favors spin canting with clockwise (counterclockwise) chirality. The inset shows the DMI vectors $\boldsymbol{D}_{ij}$ (blue vectors) between the nearest neighbors of Mn atoms (red balls). For clarity, the chalcogen atoms are omitted.

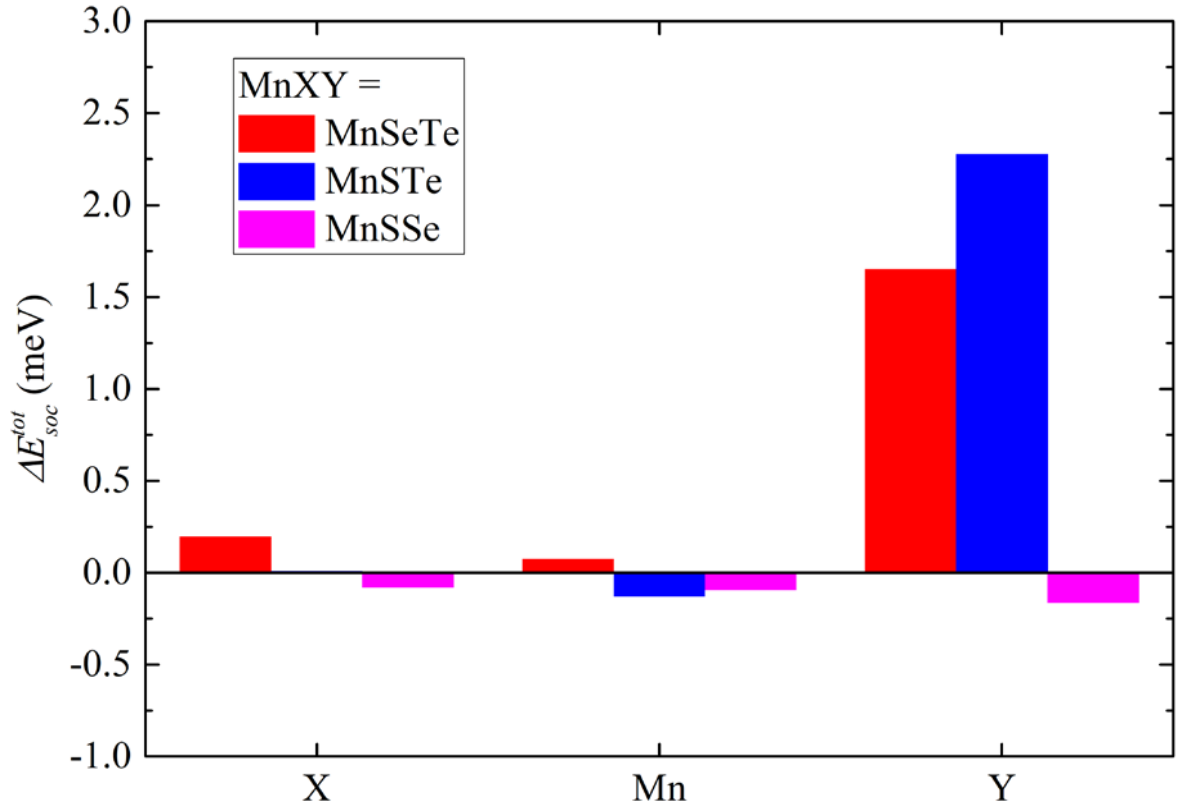

**Figure 3.** Atomic-layer-resolved localization of the SOC energy $\Delta E_{soc}$ for the Janus MnXY monolayers. As can be seen $\Delta E_{soc}$ is dominated by heavy chalcogen atom Y.

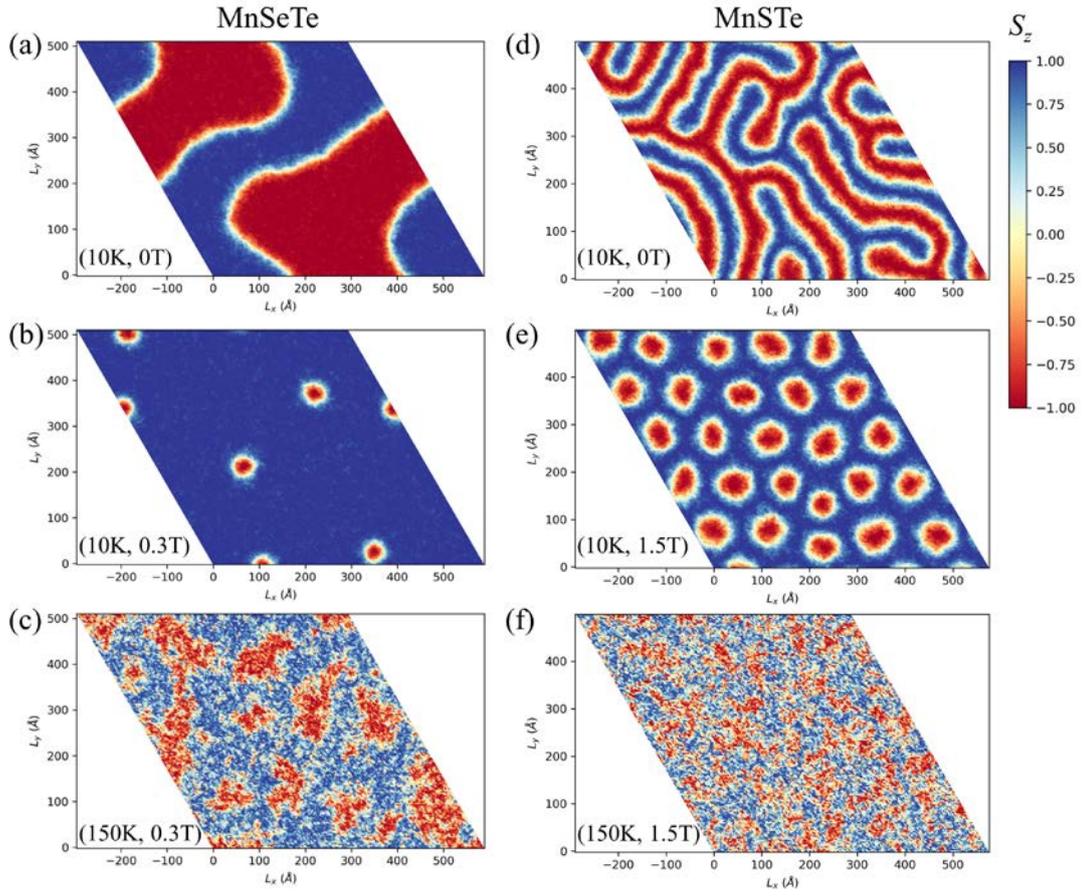

**Figure 4.** Spin textures for (a)-(c) MnSeTe and (d)-(f) MnSTe monolayers in real space. The corresponding temperatures and external fields for the simulations are labeled inside each figures. The color map indicates the out-of-plane spin component of Mn atoms.

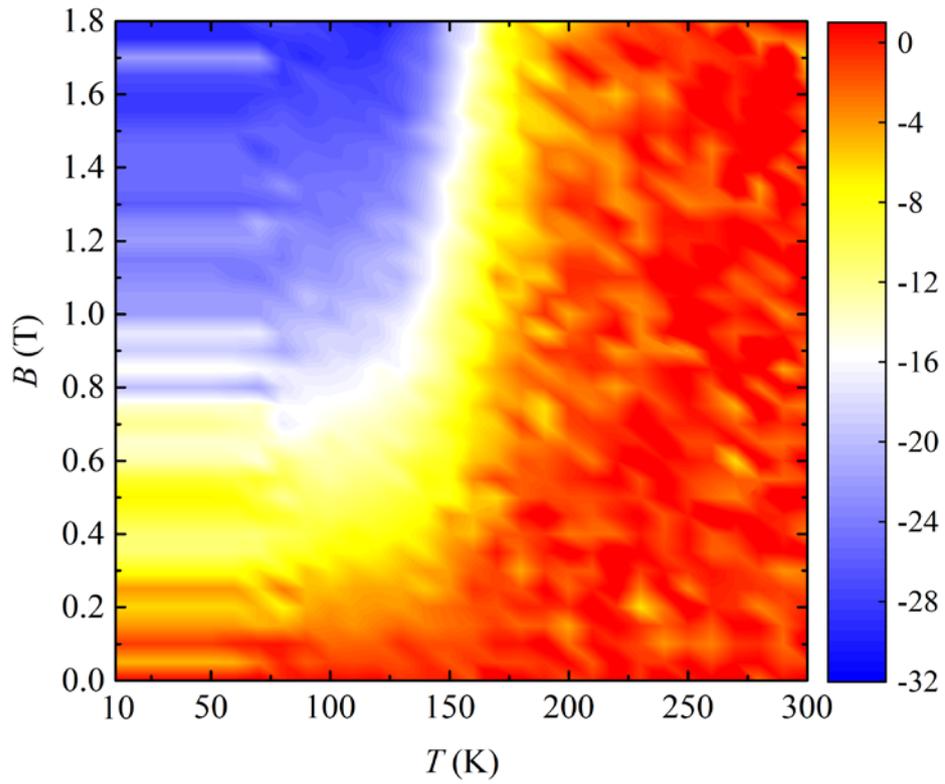

**Figure 5.** The topological charge $Q$ per supercell of MnSTe monolayers as a function of temperature and external magnetic field, calculated from MC simulations.